\documentclass[%
 preprint,
 amsmath,amssymb,
 aps, physrev,
showkeys,
]{revtex4-2}
\usepackage{multirow}
\usepackage{mathrsfs}
\usepackage{tabularx}
\usepackage{booktabs} 
\usepackage{braket}
\usepackage{graphicx}% Include figure files
\usepackage{dcolumn}% Align table columns on decimal point
\usepackage{bm}% bold math
\begin{document}

\preprint{APS/123-QED}

\title{\textbf{Quantum Direct Steganography Scheme Based on Modified Generator Projection Directions of Steane Code over a Single-Type Pauli Channel} 
}% 

\author{Chaolong Hao}
 % \altaffiliation[]{School of Information Systems Engineering, Information Engineering University}%Lines break automatically or can be forced with \\
\author{Quangong Ma}%
 \email{Contact author: quangongma@163.com}
\affiliation{%
 % Authors' affiliations\\
 %  Include all institutions where the work was conducted: department or division, institution, city, state (if relevant), and country, in this order.
 School of Information Systems Engineering, Information Engineering University, 62 Science Avenue, Zhengzhou, 450001, China
}%

\author{Dan Qu}
\email{Contact author: qudan\_xd@163.com}
\affiliation{
School of Information Systems Engineering, Information Engineering University, 62 Science Avenue, Zhengzhou, 450001, China
}%
\affiliation{
 Laboratory for Advanced Computing and Intelligence Engineering, Wuxi 214000, Jiangsu, China
}%
\author{Dawei Shi}
\affiliation{%
 Laboratory for Advanced Computing and Intelligence Engineering, Wuxi 214000, Jiangsu, China
}%

% \collaboration{CLEO Collaboration}%\noaffiliation

\date{\today}% It is always \today, today,

\begin{abstract}
In quantum mechanics, measurements of a quantum state in various directions yield distinct outcomes, a principle that forms the foundation of quantum communication theory. This paper expands upon this concept by introducing a method to modify generator projection directions (MGPD) within quantum stabilizer codes. Employing the Steane code ($(7,1,3)$ code), as a fundamental carrier, we develop a novel scheme for direct quantum steganography across a single-type Pauli channel. The infeasibility of eavesdropping decoding under MGPD is proven. We detail the steganographic encoding and decoding schemes, corresponding quantum circuits, and eavesdropping detection principles. We also use a 'Sudoku'-style strategy to balance steganographic error probabilities and provide the complete steganography protocol. Relative to existing studies, the MGPD method achieves embedding rates approaching and attaining the upper limit of the information capacity for the $(n,k,d)=(7,1,3)$ code within a noise probability range of approximately $1/(n+1)=12.5\%$. It also reduces the consumption of auxiliary keys from $O(\log{(N)})$ to $O(1)$, while enabling eavesdropping detection and steganography of arbitrary quantum states. We investigate its potential applications in quantum communication and assess its benefits in the context of secret information transmission and eavesdropping detection in noisy channels. Although the MGPD method incorporates certain idealized assumptions and limitations, it provides novel perspectives on the concealment of quantum information.
% \begin{description}
% Secondary publications and information retrieval purposes.
% You may use the \texttt{description} environment to structure your abstract;
% use the optional argument of the \verb+\item+ command to give the category of each item. 
% \end{description}
\end{abstract}

\keywords{Quantum steganography, Stabilizer code, Modified generator projection directions (MGPD), Pauli channel, Quantum communication}%Use showkeys class option if keyword
\maketitle

%\tableofcontents
% \section{\label{sec:level1}First-level heading:\protect\\ The line
% break was forced \lowercase{via} \textbackslash\textbackslash}
\section{\label{section1}Introduction}
Steganography, the art of embedding information within a carrier to evade detection by unauthorized individuals, has undergone significant evolution. Initially employing ancient invisible inks, the technique has advanced to incorporate modern digital and network-based methods \cite{ref1}. This technology, along with watermarking, has been widely implemented across various sectors, including military intelligence, copyright protection, and covert communication, due to technological progress \cite{ref2,ref3}. Since the 1980s, the rapid expansion of quantum information science has led to the emergence of technologies like Quantum Key Distribution (QKD) \cite{ref4,ref5,ref6,ref7} and Quantum Secure Direct Communication (QSDC) \cite{ref8,ref9,ref10}. These technologies offer unconditional security, which is superior to that of traditional cryptographic methods. In parallel, research into quantum-based information hiding, utilizing properties such as superposition and entanglement \cite{ref11}, has attracted considerable research interest.

From the standpoint of steganographic carriers, quantum information hiding techniques can be classified into three distinct categories: 1) Channel and Coding Methods: These approaches camouflage secret messages by mimicking the statistical distribution of noise inherent to the physical channel and embed them within Quantum Error Correction Codes (QECC) for secure transmission \cite{ref12,ref13,ref14,ref15,ref16,ref17}. 2) State Exchange and Protocol Methods: These techniques involve subtle modifications to communication protocols and exploit methods such as quantum teleportation to create covert communication channels \cite{ref18,ref19,ref20,ref21}. 3) Media Information Methods: These methods encode images or audio into quantum states and apply classical steganographic techniques such as the Least Significant Bit (LSB) method, further enhancing performance with algorithms like quantum Fourier transforms, random walks, and matrix coding \cite{ref22,ref23,ref24,ref25,ref26,ref27,ref28}. Among these categories, channel and coding methods serve as the foundational layer. To enhance resistance to noise, some techniques in the latter two categories also integrate QECC \cite{ref20,ref28}.

\begin{figure}
\centering
\includegraphics[width=0.5\textwidth]{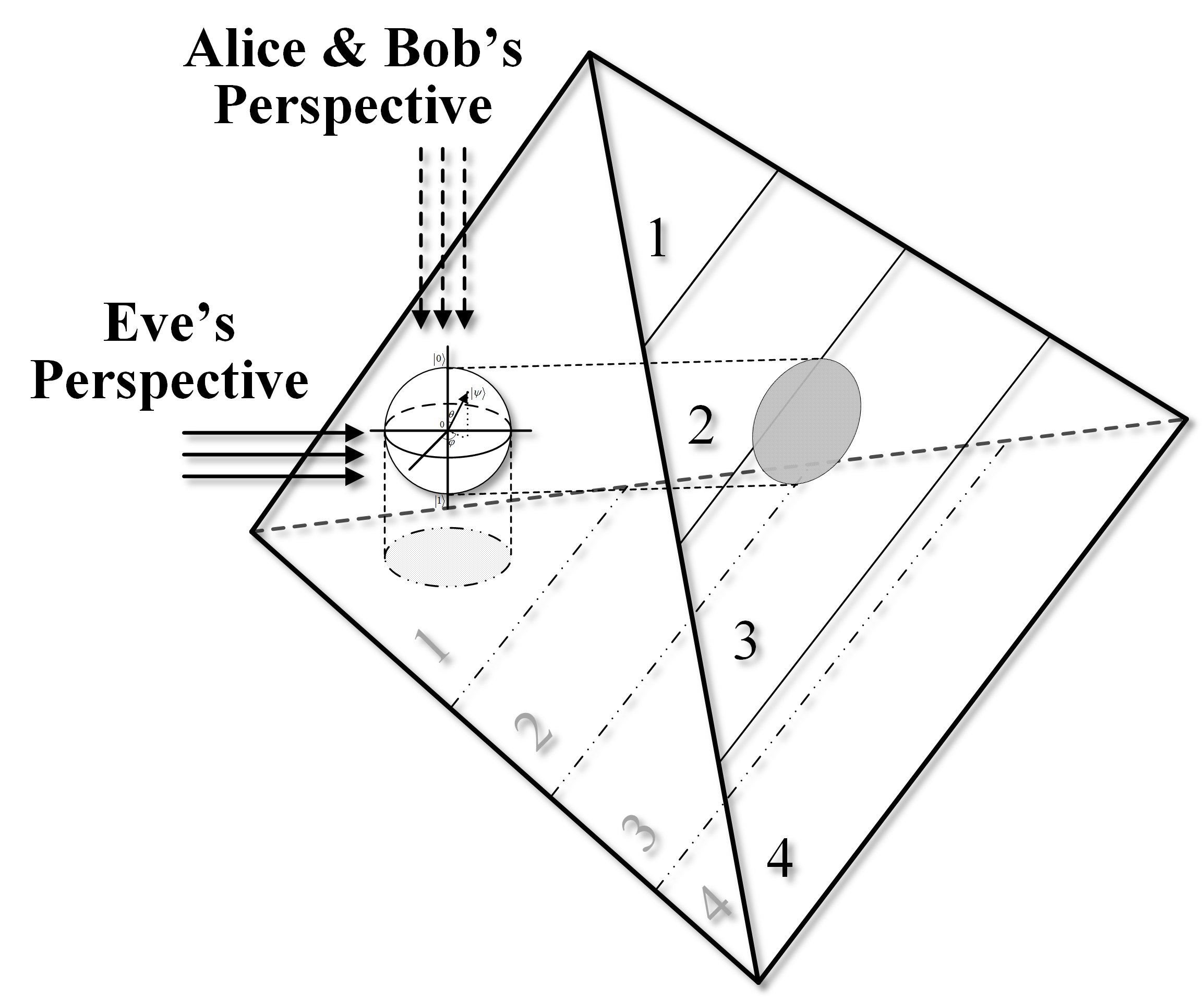}
\caption{Schematic diagram of quantum state projections from various perspectives. The tetrahedron's faces represent different directions of error-correcting code encoding, while the regions labeled 1-4 denote error categories.}
\label{figure1}
\end{figure}

Quantum theory pioneer Werner Heisenberg articulated the observer effect in quantum mechanics with the statement, 'the very act of observing disturbs the system.' This principle is fundamental to protocols such as Quantum Key Distribution (QKD) and Quantum Secure Direct Communication (QSDC). Eavesdroppers, due to incorrect measurement directions, not only fail to acquire complete and accurate information but also disturb the system, which reveals their eavesdropping activities. Applying this concept to quantum steganography, if Alice and Bob utilize measurement directions that differ from Eve's, they will perceive the quantum system differently, as illustrated in Fig. \ref{figure1}. By employing specific strategies, Alice and Bob can potentially transmit secret quantum information securely and covertly, thereby complicating Eve's ability to detect their communication.

Based on the above insights, this paper investigates a quantum steganography scheme a QECC framework. We use the Steane code as the base carrier and modify the projection directions of the encoding generators to directly transmit secret quantum states. The main contributions of this paper are: 1) We prove the theoretical infeasibility of Eve’s error correction and decoding under different projection measurement directions using the necessary and sufficient conditions for stabilizer code correction. 2) For a single-type Pauli channel, we propose the specific implementation of the MGPD method and the corresponding quantum circuits, and introduce a 'Sudoku'-style direction selection strategy to balance channel error probabilities, making the steganographic behavior resistant to Eve’s statistical analysis. 3) We construct the MGPD steganography protocol, which offers advantages over previous quantum steganography schemes based on QECC in terms of embedding rate, auxiliary key consumption, eavesdropping detection capability, and the form of secret information. 4) We explore the potential of our scheme for tasks such as QKD and QSDC, analyzing its advantages in secret information form and eavesdropping detection in noisy channels.

The organization of this paper is outlined as follows: Sect. \ref{section2} presents the foundational concepts of quantum steganography, Pauli channels, and stabilizer codes, and establishes the proposition that misaligned directions preclude Eve from accurately decoding the information. Sect. \ref{section3} elaborates on the execution of the MGPD steganography scheme, encompassing quantum circuit designs, principles of eavesdropping detection, strategies for balancing probabilities, and the overall protocol workflow. Sect. \ref{section4} conducts an analysis of the MGPD scheme and compares it with existing research in the realms of information concealment and secure communication. Sect. \ref{section5} concludes the paper with a summary of the findings.

\section{\label{section2}Preliminary}

\subsection{\label{section2.1}Basic Formalism of Quantum Steganography}

\begin{figure}
\centering
\includegraphics[width=1\textwidth]{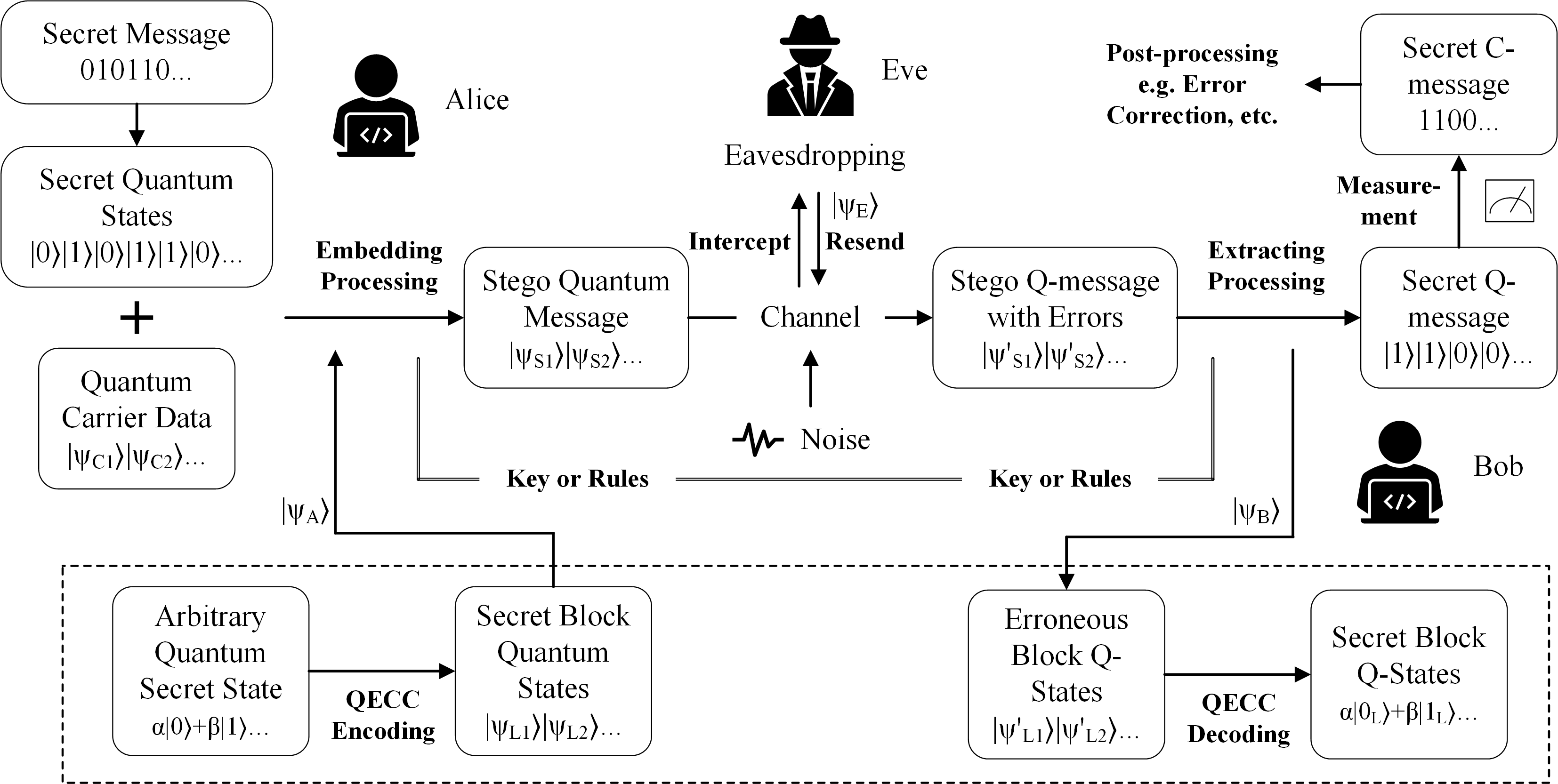}
\caption{Schematic diagram of the quantum steganography scheme.}
\label{figure2}
\end{figure}

Steganography serves as a principal technique for concealing information, applicable to both spatial and temporal transmissions (e.g., Alice embeds secret information within a quantum register for Bob to retrieve at a later time). The primary objectives of quantum steganography can be summarized as follows: 1) Communication: Alice must be capable of conveying classical or quantum information to Bob. 2) Secrecy: The embedded secret message within the cover data should be effectively concealed, rendering it challenging for Eve to discern. Furthermore, the capacity and robustness are pivotal criteria for assessing the efficacy of steganography schemes. Fig. \ref{figure2} depicts the general process of quantum steganography, where the area outside the dashed box signifies the standard procedure, and the area within the dashed box highlights the encoding and decoding steps of the direct quantum state steganography method introduced in this paper.

\subsection{\label{section2.2}Pauli Channels}
Pauli operators constitute a fundamental set of operators within quantum theory, comprising $2\times2$ unitary Hermitian matrices that satisfy anticommutation relations, as shown in Eq. \ref{equation1} and \ref{equation2}.

\begin{equation}
X=
    \begin{bmatrix}
    0 & 1  \\
    1 & 0  \\
    \end{bmatrix},
Y=
    \begin{bmatrix}
    0 & -i  \\
    i & 0  \\
    \end{bmatrix}
    ,
Z=
    \begin{bmatrix}
    1 & 0  \\
    0 & -1  \\
    \end{bmatrix}
    ,
\label{equation1}
\end{equation}

\begin{equation}
\{X,Y\}=\{Y,Z\}=\{Z,X\}=0,
\label{equation2}
\end{equation}
where \{a,b\}=$ab+ba$.

The bit-flip channel (BC) is a quantum noise model where a qubit has a probability $p$ of undergoing a bit-flip error, which changes the state from $\ket{0}$ to $\ket{1}$ or vice versa. Mathematically, the channel can be described as:

\begin{equation}
\mathcal{E}(\rho )=(1-p)\rho +pX\rho X
\label{equation3}
\end{equation}

Similarly, the mathematical descriptions of the phase-flip channel (PC) and the bit-phase flip channel (BPC) are given by Eq. \ref{equation4} and \ref{equation5}, respectively.

\begin{equation}
\mathcal{E}(\rho )=(1-p)\rho +pZ\rho Z
\label{equation4}
\end{equation}

\begin{equation}
\mathcal{E}(\rho )=(1-p)\rho +pY\rho Y
\label{equation5}
\end{equation}

Quantum channels are not merely mathematical constructs; they also represent actual physical phenomena. For instance, variations in external electromagnetic fields can induce bit-flip errors in superconducting qubits. Minor alterations in temperature and magnetic fields may result in phase-flip errors within ion trap systems. Additionally, high-energy particle collisions can precipitate both bit and phase flip errors in diamond NV center systems.

\subsection{\label{section2.3}Stabilizer Formalism}

The stabilizer formalism, introduced by Gottesman \cite{ref29}, has been instrumental in the advancement of quantum error correction theory. Stabilizers are derived from the Pauli group. The Pauli group $G_1$ for a single qubit is defined as $G_1\equiv\alpha\{I,X,Y,Z\},\alpha=\{\pm1,\pm{i}\}$. It is clear that $G_1$ is an Abelian group. The Pauli group on n-qubits is defined as $G_m\equiv{G_1^{\otimes{n}}}$.

A stabilizer $S$ is defined as a subgroup of $G_n$, typically expressed in terms of its generators: $S\equiv\langle{g_1,g_2,...,g_l}\rangle$, satisfying $\forall{k},g_k^2=I$, all mutually commute, i.e. $\forall{j,k},[g_j,g_k]=g_jg_k-g_kg_j=0$. The vector space $V_s$ is defined as the set of states stabilized by all elements of $S$, meaning that for all $\forall\ket{\psi}\in V_s, \prod_{k}g_k\ket{\psi}=\ket{\psi}$.

\subsection{\label{section2.4}Modified Generator Projection Direction Encoding}

As discussed in Sect. \ref{section2.1}, the objective of steganography is to enable Alice and Bob to transmit secret information accurately while making it challenging for Eve to detect and extract that information. To accomplish this, we propose to alter the stabilizer encoding by adjusting the generator projection direction, thereby hindering Eve's ability to correctly detect and perform error correction without knowledge of the specific rule.

Let us now consider the general conditions for quantum error correction (refer to \cite{ref11}, Theorem 10.1). Let $P$ be the projector onto a quantum code, and let $\{E_i\}$ denote the set of error operators that describe noisy quantum operations. The necessary and sufficient condition for the correctability of errors is that:

\begin{equation}
PE_i^{\dagger}E_jP=\alpha_{ij}P,
\label{equation6}
\end{equation}

for some Hermitian matrix $\alpha$ of complex numbers.

Within the framework of stabilizer encoding, the projector $P$ is represented as: $P=(1/2^l)\prod_{k=1}^l(I+g_k)$, where $l$ denotes the number of generators $g_k$. We propose to modify $P$ by altering the projection direction orthogonally for some of its dimensions, while leaving the other dimensions unaltered. For instance, $\tilde{P}=(1/2^l)(I-g_h)\prod_{k\neq{h}}(I+g_k)$.

The subsequent proposition demonstrates that in this context, Eve, being uninformed of the modification rule, will be unable to correct errors accurately if she persists in decoding based on the original projection directions. For the sake of simplicity, we consider the modification of a single generator direction; the extension to multiple directions is analogous.

\textbf{\textit{Proposition 1}}: It is not possible to find a set of errors $\{E_i\}$ that the stabilizer code, defined by the projection operator $P=(1/2^l)\prod_{k=1}^l(I+g_k)$ can correct, such that $\tilde{P}E_i^{\dagger}E_j\tilde{P}=\beta_{ij}P$, where $\tilde{P}=(1/2^l)(I-g_h)\prod_{k\neq{h}}(I+g_k)$, and $\beta_{ij}$ constitutes a Hermitian matrix.

In this context, $\tilde{P}$ can be interpreted as the projection encoding operation following a directional modification, as perceived by Alice and Bob, whereas $P$ represents the projection operation in the original direction, which Eve attempts to use for error correction.

\textbf{\textit{Proof}}: 

Let $\bar{P_h}=(1/2^{l-1})\prod_{k\neq{h}}(I+g_k)$. It is straightforward to see from the expressions of $P$ and $\tilde{P}$ that:

\begin{equation}
\tilde{P}=P-g_h\left(\frac{1}{2^{l-1}}\prod_{k\neq{h}}(I+g_k)\right)=P-g_h\bar{P_h}.
\label{equation7}
\end{equation}

Combining Eq. \ref{equation6}, direct calculation yields:

% \begin{small}
% \begin{equation}

\begin{align}
&\tilde{P}E_i^{\dagger}E_j\tilde{P}=(P-g_h\bar{P_h})E_i^{\dagger}E_j(P-g_h\bar{P_h})\notag\\
&=PE_i^{\dagger}E_jP-PE_i^{\dagger}E_jg_h\bar{P_h}-g_h\bar{P_h}E_i^{\dagger}E_jP+g_h\bar{P_h}E_i^{\dagger}E_jg_h\bar{P_h}\notag\\
&=\alpha_{ij}P+D_{i,j,h},
\label{equation8}
\end{align}
% \end{equation}
% \end{small}
where,

\begin{equation}
D_{i,j,h}=-PE_i^{\dagger}E_jg_h\bar{P_h}-g_h\bar{P_h}E_i^{\dagger}E_jP+g_h\bar{P_h}E_i^{\dagger}E_jg_h\bar{P_h}
\label{equation9}
\end{equation}
For $D_{i,j,h}$, we discuss the problem in three cases.

\textbf{Case1}: $E_i^{\dagger}E_j$ commutes with all the generators $g_k$. That is, $[E_i^{\dagger}E_j,g_k]=0,k=1,2,...l$. In this situation,

\begin{align}
&D_{i,j,h}=-2g_h\bar{P_h}PE_i^{\dagger}E_j+\bar{P_h}E_i^{\dagger}E_j\notag\\
&=\left(-2g_h\bar{P_h}\left(\frac{1}{2}(I+g_h)\bar{P_h}\right)+\bar{P_h}\right)E_i^{\dagger}E_j\notag\\
&=-g_h\bar{P_h}E_i^{\dagger}E_j=(\tilde{P}-P)E_i^{\dagger}E_j.
\label{equation10}
\end{align}

Substituting Eq. \ref{equation10} into \ref{equation8}, we get:
\begin{equation}
\tilde{P}E_i^{\dagger}E_j\tilde{P}=\alpha_{ij}P+(\tilde{P}-P)E_i^{\dagger}E_j.
\label{equation11}
\end{equation}

Due to the orthogonality of $P$ and $\tilde{P}$, it is clearly shown that there does not exist $\beta_{ij}$ satisfying the conditions.

\textbf{Case2}: $E_i^{\dagger}E_j$ only anticommutes with $g_h$. That is, $\left\{E_i^{\dagger}E_j,g_h=0\right\}$, while $[E_i^{\dagger}E_j,g_k]=0,k\neq{h}$. In this situation,

\begin{align}
&D_{i,j,h}=-\left(\frac{1}{2}(I+g_h)\bar{P_h}\right)g_h\bar{P_h}E_i^{\dagger}E_j-g_h\bar{P_h}\left(\frac{1}{2}(I-g_h)\bar{P_h}\right)E_i^{\dagger}E_j-g_h\bar{P_h}g_h\bar{P_h}E_i^{\dagger}E_j\notag\\
&=\left(-\frac{1}{2}(I+g_h)-\frac{1}{2}(g_h-I)-I\right)\bar{P_h}E_i^{\dagger}E_j\notag\\
&=-(I+g_h)\bar{P_h}E_i^{\dagger}E_j=-2PE_i^{\dagger}E_j.
\label{equation12}
\end{align}

Then, $\tilde{P}E_i^{\dagger}E_j\tilde{P}=\alpha_{ij}P-2PE_i^{\dagger}E_j$, indicating that there does not exist $\beta_{ij}$.

\textbf{Case3}: $E_i^{\dagger}E_j$ anticommutes with at least one generator other than $g_h$. Without loss of generality, suppose $\left\{E_i^{\dagger}E_j,g_1=0\right\}$. Due to 

\begin{equation}
E_i^{\dagger}E_j\bar{P_h}=(I-g_1)(\frac{1}{2^{l-1}}\prod_{k\neq{1}}(I+g_k))E_i^{\dagger}E_j,
\label{equation13}
\end{equation}

and $\bar{P_h}(I-g_1)=0$, therefore, all three terms of the polynomial on the right-hand side of Eq. \ref{equation9} are zero, i.e. $D_{i,j,h}=0$. Let $\beta_{ij}=\alpha_{ij}$, the given conditions are satisfied. Combining the above three cases, \textbf{\textit{Proposition 1}} is proved.

\section{\label{section3}Quantum Steganography with MGPD of Steane Code}

In the preceding section, we discussed the pertinent concepts of quantum information steganography and established the viability of the steganographic scheme presented in this paper. This section will utilize the $(7,1,3)$ Steane code \cite{ref30} to devise suitable strategies and formulate the steganography protocol, prioritizing the dual goals of communication and secrecy for the steganographic message.

The $(7,1,3)$ Steane code is a renowned quantum error-correcting code (QECC) that encodes a single logical qubit into seven physical qubits. It is capable of correcting up to one arbitrary quantum error, which may manifest as a bit-flip error $X$, a phase-flip error $Z$, or a combination of both (i.e. $Y$). 

Due to its practical implementation advantages, this code is frequently selected for error correction across various channels. The stabilizer generators for the Steane code are detailed in Table \ref{table1}. During the error correction phase, the syndrome is determined through measurements of the generators, which facilitates the identification of the error and the application of the corresponding operation to restore the logical state.

\begin{table}
\setlength{\tabcolsep}{25pt}
  \centering
  \caption{The generators for the $(7,1,3)$ code. Each term represents an error on the corresponding qubit. For example, $X_4,X_5,X_6,X_7=I\otimes I\otimes I\otimes X \otimes X \otimes X \otimes X$.}
% \scalebox{1.2}
    \begin{tabular}{cc}
\toprule%第一道横线
    Name& Operator\\  
\midrule%第二道横线
    $g_1$ & $X_4,X_5,X_6,X_7$\\
    $g_2$ & $X_2,X_3,X_6,X_7$\\
    $g_3$ & $X_1,X_3,X_5,X_7$\\
    $g_4$ & $Z_4,Z_5,Z_5,Z_7$\\
    $g_5$ & $Z_2,Z_3,Z_6,Z_7$\\
    $g_6$ & $Z_1,Z_3,Z_5,Z_7$\\
    \bottomrule%第三道横线
  \end{tabular}
  \label{table1}
\end{table}

\subsection{\label{section3.1}Stego States’ Encoding and Error Correction}

We begin with two foundational assumptions: first, that Eve is aware of the generator (the original projection directions) of the $(7,1,3)$ code; second, that only single-qubit errors affect the encoded states. These assumptions are justifiable, as the generators of the $(7,1,3)$ code can be standardized for computational or communicative purposes. In scenarios where multiple qubits incur errors, concatenated coding can be implemented, re-encoding each qubit using the $(7,1,3)$ code. Thus, for our study, considering only single-qubit errors is adequate. Furthermore, our study focuses on single-type error channels (i.e. BC, PC, BPC mentioned above), which will be elaborated in the following.

As presented in Table \ref{table1}, the interaction—either commutation or anticommutation—between single-qubit error operators and the generators are detailed in Table \ref{table2}.

\begin{table}
\setlength{\tabcolsep}{4pt}
  \centering
  \caption{The relationships between the single qubit error and the generators of $(7,1,3)$ code. Here, +1 represents that the two operators commute, and -1 represents that they anticommute.}
% \scalebox{1.2}
    \begin{tabular}{c|cccccc|c|cccccc|c|cccccc}
\toprule%第一道横线
     & $g_1$ & $g_2$ & $g_3$ & $g_4$ & $g_5$& $g_6$ & &$g_1$ & $g_2$ & $g_3$ & $g_4$ & $g_5$& $g_6$& &$g_1$ & $g_2$ & $g_3$ & $g_4$ & $g_5$& $g_6$\\  
\midrule%第二道横线
    $I$ & +1& +1& +1& +1& +1& +1& $I$ & +1& +1& +1& +1& +1& +1& $I$ & +1& +1& +1& +1& +1& +1 \\
    $X_1$ & +1& +1& +1& +1& +1& -1&$Z_1$& +1& +1& -1& +1& +1& +1& $Y_1$ & +1& +1& -1& +1& +1& -1 \\
    $X_2$ & +1& +1& +1& +1& -1& +1&$Z_2$& +1& -1& +1& +1& +1& +1& $Y_2$ & +1& -1& +1& +1& -1& +1\\
    $X_3$ & +1& +1& +1& +1& -1& -1&$Z_3$& +1& -1& -1& +1& +1& +1& $Y_3$ & +1& -1& -1& +1& -1& -1\\
    $X_4$ & +1& +1& +1& -1& +1& +1&$Z_4$& -1& +1& +1& +1& +1& +1& $Y_4$ & -1& +1& +1& -1& +1& +1\\
    $X_5$ & +1& +1& +1& -1& +1& -1&$Z_5$& -1& +1& -1& +1& +1& +1& $Y_5$ & -1& +1& -1& -1& +1& -1\\
    $X_6$ & +1& +1& +1& -1& -1& +1&$Z_6$& -1& -1& +1& +1& +1& +1& $Y_6$ & -1& -1& +1& -1& -1& +1\\
    $X_7$ & +1& +1& +1& -1& -1& -1&$Z_7$& -1& -1& -1& +1& +1& +1& $Y_7$ & -1& -1& -1& -1& -1& -1\\
    \bottomrule%第三道横线
  \end{tabular}
  \label{table2}
\end{table}

Now, let us concentrate on the projection direction along $g_1$ to demonstrate the alterations in the encoding and error correction processes when the projection is altered. At the sender's end, the standard encoding operator is given by: $P=1/{2^6}\prod_{k=1}^{6}(I+g_k)$, while Alice's modified encoding operator is: $\tilde{P_1}=1/{2^6}(I-g_1)\prod_{k=2}^{6}(I+g_k)$. At the receiver's end, Bob employs the modified generator directions for detection and correction, whereas Eve relies on the original projections. We examine two error scenarios: $E_i^{\dagger}E_j=I$ (i.e. $E_i=E_j$) and $E_i^{\dagger}E_j=Z_4$. Since $I$ commutes with all generators, it corresponds to \textbf{Case 1} in \textbf{\textit{Proposition 1}}. $Z_4$ only anticommutes with $g_1$, corresponding to \textbf{Case 2}. Consequently, neither scenario permits accurate error correction, implying that Eve cannot accurately ascertain the secret information.

\begin{figure}
\centering
\includegraphics[width=1\textwidth]{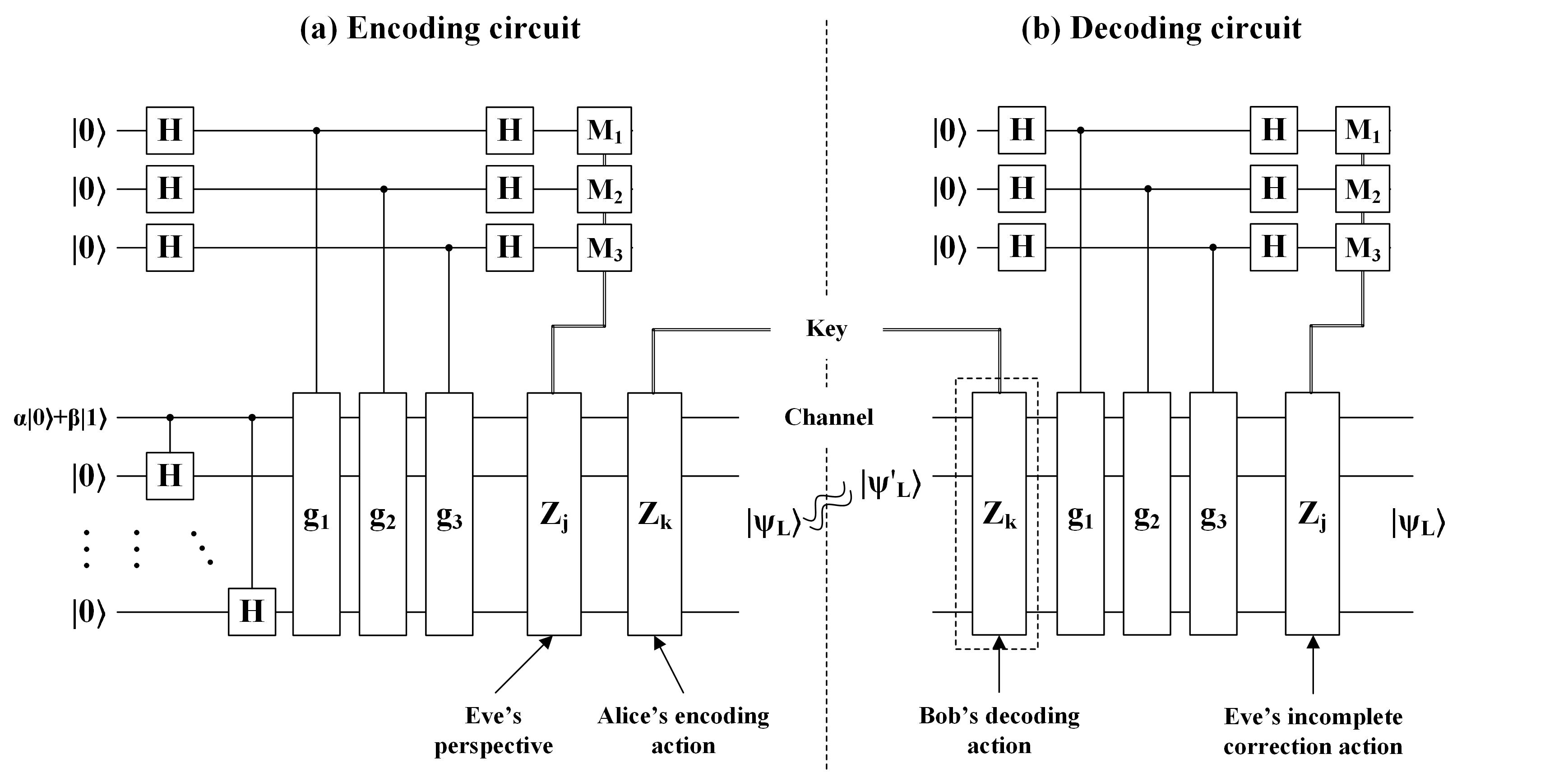}
\caption{Circuit for stego states based on the Steane Code (with $Z$-type errors as an example). Subfigure (a) illustrates the encoding circuit, while subfigure (b) depicts the decoding circuit. The variables $M_1$, $M_2$, and $M_3$ correspond to the measurement outcomes of the ancillary qubits.}
\label{figure3}
\end{figure}

The one implementation method of quantum circuits is illustrated in Fig. \ref{figure3}. We adhere to the method outlined in Reference \cite{ref31} and incorporate error operation modules, determined by commutation relations, into the standard encoding circuit. The key components include the $Z_j$ and $Z_k$ modules; specifically, the $Z_k$ module, enclosed within the dashed box, is used exclusively by Bob during the decoding phase, provided he is cognizant of the protocol. Alice and Bob control the $Z_k$ module to modify the direction of $g_1$, with additional details elaborated in Eqs. \ref{equation17} to \ref{equation19} within Sect. \ref{section3.2}.

In Fig. \ref{figure3}, the key modules for encoding and error detection are mathematically represented as follows:

\begin{align}
&(H\otimes I)\dot{g}(H\otimes I)\ket{0}\otimes\ket{\psi}\notag\\
&=(\ket{+}\bra{+}\otimes I+\ket{-}\bra{-}\otimes g)\ket{0}\otimes\ket{\psi}\notag\\
&=\frac{1}{\sqrt2}\ket{0}\otimes(I+g)\ket{\psi}+\frac{1}{\sqrt2}\ket{1}\otimes(I-g)\ket{\psi}.
\label{equation14}
\end{align}

During the detection phase, generators are measured in isolation. The measurement outcome is 0 if the error commutes with a generator, and 1 if it anticommutes. By consulting Table \ref{table2}, the type of error can be identified and subsequently corrected. For instance, if a $Z_6$ error occurs, it anticommutes with $g_1$ and $g_2$, while commuting with the remaining generators, thus:

\begin{align}
&\ket{0}\otimes(I+g_1)Z_6\prod_{k=1}^6(I+g_k)\ket{\psi}\notag\\
&=\ket{0}\otimes(I+g_1)(I-g_1)(I-g_2)\prod_{k=3}^6(I+g_k)Z_6\ket{\psi}=0,\notag\\
&\ket{1}\otimes(I+g_1)Z_6(I-g_1)\prod_{k=2}^6(I+g_k)\ket{\psi}\ne0.
\label{equation15}
\end{align}

The measurement outcomes 1 for $g_1$.

The preceding scenario pertains to the use of the original projection direction. However, with Alice's encoding projection direction now aligned with $-g_1$, the measurement outcome for the $Z_6$ error on $g_1$ is the converse of what it was in the original direction (refer to Eq. \ref{equation16}). The measurement of $g_1$ yields 0.

\begin{align}
&\ket{0}\otimes(I+g_1)Z_6(I-g_1)\prod_{k=2}^6(I+g_k)\ket{\psi}\notag\\
&=\ket{0}\otimes(I+g_1)(I+g_1)(I-g_2)\prod_{k=2}^6(I+g_k)Z_6\ket{\psi}\ne0,\notag\\
&\ket{1}\otimes(I-g_1)Z_6(I-g_1)\prod_{k=2}^6(I+g_k)\ket{\psi}=0.
\label{equation16}
\end{align}

Consequently, the syndromes measured by Eve for the generators $g_1$ through $g_6$ shift from 110000 to 010000. Based on Table \ref{table2}, Eve incorrectly identifies a $Z_2$ error rather than the $Z_6$ error. Table \ref{table3} displays the discrepancy between the actual errors and Eve's perceived errors. While Eve's measurements are entirely erroneous, Bob, who is cognizant of the protocol, can use Table \ref{table3} to adjust and rectify the errors. This process ensures the accurate transmission of the quantum state from Alice to Bob, fulfilling one of the primary objectives of steganography.

\begin{table}
\setlength{\tabcolsep}{2.5pt}
  \centering
  \caption{Error analysis from Eve's perspective versus actual errors post-modification of the $g_1$ direction.}
% \scalebox{1.2}
    \begin{tabular}{ccccccccc}
\toprule%第一道横线
    Eve's syndrome& 100000& 101000& 110000& 111000& 000000& 001000& 010000& 011000\\  
% \midrule%第二道横线
    Real error& $I$ & $Z_1$& $Z_2$& $Z_3$& $Z_4$& $Z_5$& $Z_6$& $Z_7$\\
    Eve’s error& $Z_4$& $Z_5$& $Z_6$& $Z_7$ & $I$ & $Z_1$& $Z_2$& $Z_3$\\
    \bottomrule%第三道横线
  \end{tabular}
  \label{table3}
\end{table}

It is crucial to acknowledge that this steganographic method is effective solely for single-Pauli error channels. Should the channel encounter multiple error types concurrently (e.g., depolarizing channels exhibiting $X$, $Y$, and $Z$ errors), this approach becomes inapplicable. For instance, altering the $g_1$ direction may result in $X$ and $Y$ errors yielding unanticipated syndromes (e.g., 100001, not present in Table \ref{table2}), prompting Eve to suspect non-single-qubit errors. Consistent occurrence of such events could arouse Eve's suspicion. Current research on steganographic methods for depolarizing channels predominantly focuses on employing a key to deliberately introduce a small number of errors that encode secret messages into the channel, as elaborated in references \cite{ref13,ref14,ref15,ref16}.

Let us examine a scenario where the eavesdropper, Eve, intercepts and measures the secret quantum state to obtain the error syndrome before resending it to Bob, akin to a measure-resend attack. The question arises whether Bob can detect this interference and still correct the errors. The answer is that while Bob can indeed detect the interference, he cannot successfully complete the error correction.

As depicted in Fig. \ref{figure2}, the encoded quantum state $\ket{\psi_{A}}$ travels from Alice to Eve and then from Eve to Bob, passing through two channels (Channel-1 and Channel-2, both with the same channel parameters). This results in the quantum state being subjected to noise twice. To evade detection, Eve must perform error correction before resending, attempting to make the state appear as though it has only encountered noise once. However, since Alice has applied the MGPD method, Eve's error correction will not function as expected.

We illustrate this with an example of modifying the $g_1$ direction, as shown in Table \ref{table4}. From Table \ref{table4}, it is evident that the quantum state  $\ket{\psi_{E}}$ resent by Eve is equivalent to applying a $Z_4$ error to the initial encoded state $\ket{\psi_{A}}$.

\begin{table}
\setlength{\tabcolsep}{10pt}
  \centering
  \caption{Error changes in state before and after eavesdropping.}
% \scalebox{1.2}
    \begin{tabular}{ccccccccc}
\toprule%第一道横线
    Errors in Channel-1& $I$ & $Z_1$& $Z_2$& $Z_3$& $Z_4$& $Z_5$& $Z_6$& $Z_7$\\  
% \midrule%第二道横线
    Eve’s action&  $Z_4$& $Z_5$& $Z_6$& $Z_7$ & $I$ & $Z_1$& $Z_2$& $Z_3$\\
    Equivalent syndrome error& $Z_4$& $Z_4$& $Z_4$& $Z_4$ & $Z_4$ & $Z_4$& $Z_4$& $Z_4$\\
   
    \bottomrule%第三道横线
  \end{tabular}
  \label{table4}
\end{table}

Assume the probability of a single physical qubit error in the channel is $p$ (Note that within a logical block consisting of 7 qubits, there is assumed to be only a single erroneous assumption). Consequently, the probability of no error occurring is
$p_0=1-7p$, and the probability of an error occurring is $p_e=7p$. The possible error scenarios for the quantum state $\ket{\psi_{B}}$ received by Bob are shown in Table \ref{table5}. Upon conducting a statistical analysis, Bob will observe two discrepancies: 1) The probability of no error $p_0^{'}=p$; 2) $p(Z_4)=1-7p\neq{p(Z_k)}=p,k\neq{4}$. These anomalies enable Bob to detect the presence of the eavesdropper, Eve, prompting him to alert Alice to cease communication. Since the two channel errors surpass the error-correcting capacity of the Steane code, Bob is unable to complete the error correction and must discard the compromised quantum states.

\begin{table}
\setlength{\tabcolsep}{10pt}
  \centering
  \caption{Equivalent errors in the syndrome of the state received by Bob.}
% \scalebox{1.2}
    \begin{tabular}{ccccccccc}
\toprule%第一道横线
    Errors in Channel-2& $I$ & $Z_1$& $Z_2$& $Z_3$& $Z_4$& $Z_5$& $Z_6$& $Z_7$\\  
% \midrule%第二道横线
    Probability&  $1-7p$& $p$& $p$& $p$ & $p$ & $p$& $p$& $p$\\
    Overlaid error& $Z_4$& $Z_4$& $Z_4$& $Z_4$ & $Z_4$ & $Z_4$& $Z_4$& $Z_4$\\
   Equivalent syndrome error& $Z_4$& $Z_5$& $Z_6$& $Z_7$ & $I$ & $Z_1$& $Z_2$& $Z_3$\\
    \bottomrule%第三道横线
  \end{tabular}
  \label{table5}
\end{table}

\subsection{\label{section3.2} Probability Adjustment Strategies to Resist Statistical Analysis}

In the method outlined in the previous subsection, Alice and Bob fulfill their communication objective by consenting to alter the encoding projection direction. This subsection focuses on an additional goal—security—and delves into techniques to bolster the imperceptibility  of the steganographic behavior. In the preceding section, Bob could detect eavesdropping based on shifts in error probabilities, and Eve could also potentially conduct steganalysis using statistical channel characteristics. For instance, if the error probability for a particular qubit within a 7-qubit encoded block deviates markedly from the rest, it might arouse Eve's suspicion of steganographic activity. We will now discuss strategies to counteract Eve's suspicions.

In specific channel environments, we assume the probability of an error occurring is $p_e$ and the probability of no error occurring is $p_o$, satisfying $p_o+p_e=1$. Typically, $p_o\neq{p_e}$. However, upon examining Table \ref{table3}, it becomes evident that, from Eve's perspective, the probability of no error occurring (denoted as 
$(I)$) is equivalent to the probability of an error occurring, with the exception of the
$Z_4$ error. The probability of a $Z_4$ error stands out as distinct from that of the other qubits. This inconsistency allows Eve to easily detect anomalies through statistical analysis.

To mitigate this issue, we must employ specific strategies to regulate the fluctuations in error probabilities resulting from the MGPD. Taking the $(7,1,3)$ code as an example in the context of a phase-flip channel (with analogous methods applicable to bit-flip and bit-phase-flip channels), we can, by referring to Table \ref{table3}, establish the correlation between the errors perceived by Eve and the actual errors for various potential direction adjustments, as detailed in Table \ref{table6}.

\begin{table}
\setlength{\tabcolsep}{10pt}
  \centering
  \caption{Errors from Eve's perspective post-directional alteration.}
% \scalebox{1.2}
    \begin{tabular}{ccccccccc}
\toprule%第一道横线
    Modified directions& \multicolumn{8}{c}{Errors from Eve’s perspective} \\  
 \midrule%第二道横线
    Original&  $I$ & $Z_1$& $Z_2$& $Z_3$& $Z_4$& $Z_5$& $Z_6$& $Z_7$\\
    $g_1$& $Z_4$& $Z_5$& $Z_6$& $Z_7$ & $I$ & $Z_1$& $Z_2$& $Z_3$\\
   $g_2$& $Z_2$& $Z_3$& $I$& $Z_1$ & $Z_6$ & $Z_7$& $Z_4$& $Z_5$\\
   $g_3$& $Z_1$& $I$& $Z_3$& $Z_2$ & $Z_5$ & $Z_4$& $Z_7$& $Z_6$\\
   $g_1,g_2$& $Z_6$& $Z_7$& $Z_4$& $Z_5$ & $Z_2$ & $Z_3$& $I$& $Z_1$\\
   $g_1,g_3$& $Z_5$& $Z_4$& $Z_7$& $Z_6$ & $Z_1$ & $I$& $Z_3$& $Z_2$\\
   $g_2,g_3$& $Z_3$& $Z_2$& $Z_1$& $I$ & $Z_7$ & $Z_6$& $Z_5$& $Z_4$\\
   $g_1,g_2,g_3$& $Z_5$& $Z_6$& $Z_5$& $Z_4$ & $Z_3$ & $Z_2$& $Z_1$& $I$\\
    \bottomrule%第三道横线
  \end{tabular}
  \label{table6}
\end{table}
Recall Fig. \ref{figure3} from Sect. \ref{section3.1}, where the operator corresponding to the encoding circuit is (See Ref. \cite{ref31}):

\begin{equation}
[I+(-1)^{M_1}g_1][I+(-1)^{M_2}g_2][I+(-1)^{M_3}g_3]
\label{equation17}
\end{equation}

To obtain the correct encoding direction $(I+g_1)(I+g_2)(I+g_3)$  (Eve’s perspective), it is necessary to apply $Z_j$ to to Eq. \ref{equation17} to make the adjustment. For example:

\begin{equation}
Z_5(I-g_1)(I+g_2)(I-g_3)=(I+g_1)(I+g_2)(I+g_3)Z_5
\label{equation18}
\end{equation}

By examining the sequence of elements in the first column of Table \ref{table6} \\(i.e., $I,Z_4,Z_2,Z_1,Z_6,Z_5,Z_3,Z_7$), the relationship between  $j$ in $Z_j$  and the measurement outcomes $M_1,M_2,M_3$ can be formulated as follows:

\begin{equation}
j=4^{M_1}+2^{M_2}+1^{M_3}
\label{equation19}
\end{equation}

The value of $k$ in $Z_k$ module can be be selected by consulting Table \ref{table6} and considering the desired direction of modification.

Furthermore, it is observed that Table \ref{table6} is a square matrix, analogous to a 'Sudoku' puzzle, where $I$, $Z_1$ to $Z_7$ appear without repetition in each column, and $I$ occupies distinct positions in each row, ensuring comprehensive coverage from the original $Z_1$ to $Z_7$ (from Eve’s perspective). It is evident that Alice can uniformly apply directional modulation encoding, thereby achieving a balance in the probability of error occurrence for each physical qubit, with the exception of no-error scenarios.

Specifically, let us assume that the acceptable probability range for a single qubit to undergo a phase-flip error in the target channel is $[p,p+\delta]$, where $p$ denotes the mean error probability and $0\leq\delta\ll1$ signifies the permissible deviation. From Eve's standpoint, this corresponds to a minor alteration in the channel's characteristics, as illustrated in Eq. \ref{equation20}.

\begin{equation}
\mathscr{N}_{p+\delta}^{PF}=[1-(p+\delta)]\rho +(p+\delta)Z\rho Z
\label{equation20}
\end{equation}

This implies that any error probability falling within this interval is regarded as typical and is not likely to be flagged as unusual by Eve. Consequently, we can determine the execution probabilities $p_g$ for the seven modified directions (one execution per encoded block) outlined in Table \ref{table6}, ensuring they align with the physical behavior of the channel, as demonstrated in Eqs. \ref{equation21} and \ref{equation22}.

\begin{align}
&p_e=(1-7p_g)p+p_g(1-7p)+6p_gp\leq p+ \delta \notag\\
&\rightarrow p_g \leq \frac{\delta}{|1-8p|}
\label{equation21}
\end{align}

In addition to the constraints on the probabilities imposed by $1-7p\geq0,p+\delta \leq 1/7+\varepsilon,0 < \varepsilon \ll1$. We can get:

\begin{equation}
\left\{
\begin{aligned}
%\nonumber
&p_g \leq \frac{\delta}{|1-8p|}\\
&\delta \leq \frac{1}{7}-p+\varepsilon\\
&p \leq \frac{1}{7},p_g \leq \frac{1}{7}\\
\end{aligned}
 \label{equation22}
\right.
\end{equation}

Once the parameters $p_g$ and $\delta$ have been established in accordance with the aforementioned criteria, the proposed scheme fulfills its fundamental security objectives and effectively counters the issue of resisting statistical analysis , as initially discussed in this subsection.

Expanding on this foundation, we address two pivotal concerns: 1) The selection of modified directions and the associated key consumption. 2) The parameter setting issue.

Regarding the first issue, based on the set of modified direction-probability correspondence pairs $\left\{(\text{Original},1-7p_g),(g_1;p_g),(g_2;p_g),...,(g_1,g_2,g_3;p_g)\right\}$, Alice can construct a sequence from the set $\left\{0,1,2,...,7\right\}$ with frequencies approximating the aforementioned probabilities.
The numbers in this sequence correspond to the row numbers in Table \ref{table6}, indicating the orthogonal adjustments of the respective generator directions. For example, if $p=0.1,\delta=0.02$, then $p_g=0.1$, the sequence could be set as $K=(0,1,2,0,3,4,0,5,6,7)$ (not unique), such that the frequency of modified directions is $0.7=7p_g$, and the frequency of unmodified directions is $0.3=1-7p_g$.

Alice then repeatedly employs this sequence for encoding and communicates it to Bob. With this sequence, Bob can accurately perform error correction and decoding. In this context, unless Eve conducts statistical analysis in frames of 10, it is challenging for her to discern the error pattern. Even if Eve analyzes in frames that are multiples of 10, the subtle probability fluctuations in error occurrences on specific physical qubits, resulting solely from the modification of encoding projection directions, are unlikely to be detected. Consequently, Alice only needs to transmit a constant-sized numerical sequence as the key to Bob, thereby achieving the communication functionality of the steganography scheme.

For the second issue, we examine the embedding rate and security of the steganography scheme. In the scheme previously described, Alice directly embeds the secret message into the code-stream for transmission. Here, the encoded states with modified projection directions are designated as stego states, which Eve cannot accurately obtain, while those without modification are considered normal states, capable of being corrected by Eve and thus not used by Alice to convey secret information. Furthermore, each set of 7-qubits block represents 1 logical qubit of quantum information. If the steganographic embedding rate is defined as the ratio of the secret message's volume to the number of physical qubits in the carrier, then the embedding rate $r=(1/7)\times 7p_g=p_g$, indicating a linear positive correlation. The security index $s$ is inversely related to the distortion rate, which corresponds to the aforementioned probability deviation $\delta$; to enhance s, a smaller $\delta$ is necessary (if modeled with a linear function, it denotes a linearly decreasing relationship). During the implementation of the steganography protocol, a trade-off must be made based on practical conditions, as shown in Eq. \ref{equation22}. By integrating Eqs. \ref{equation22} and \ref{equation23}, suitable parameters $\delta$ and $p_g$ can be determined.

\begin{equation}
\left\{
\begin{aligned}
%\nonumber
&r \propto \delta\\
&s \propto -\delta\\
\end{aligned}
 \label{equation23}
\right.
\end{equation}

\subsection{\label{section3.3}Steganography Protocol}
\begin{figure}
\centering
\includegraphics[width=0.5\textwidth]{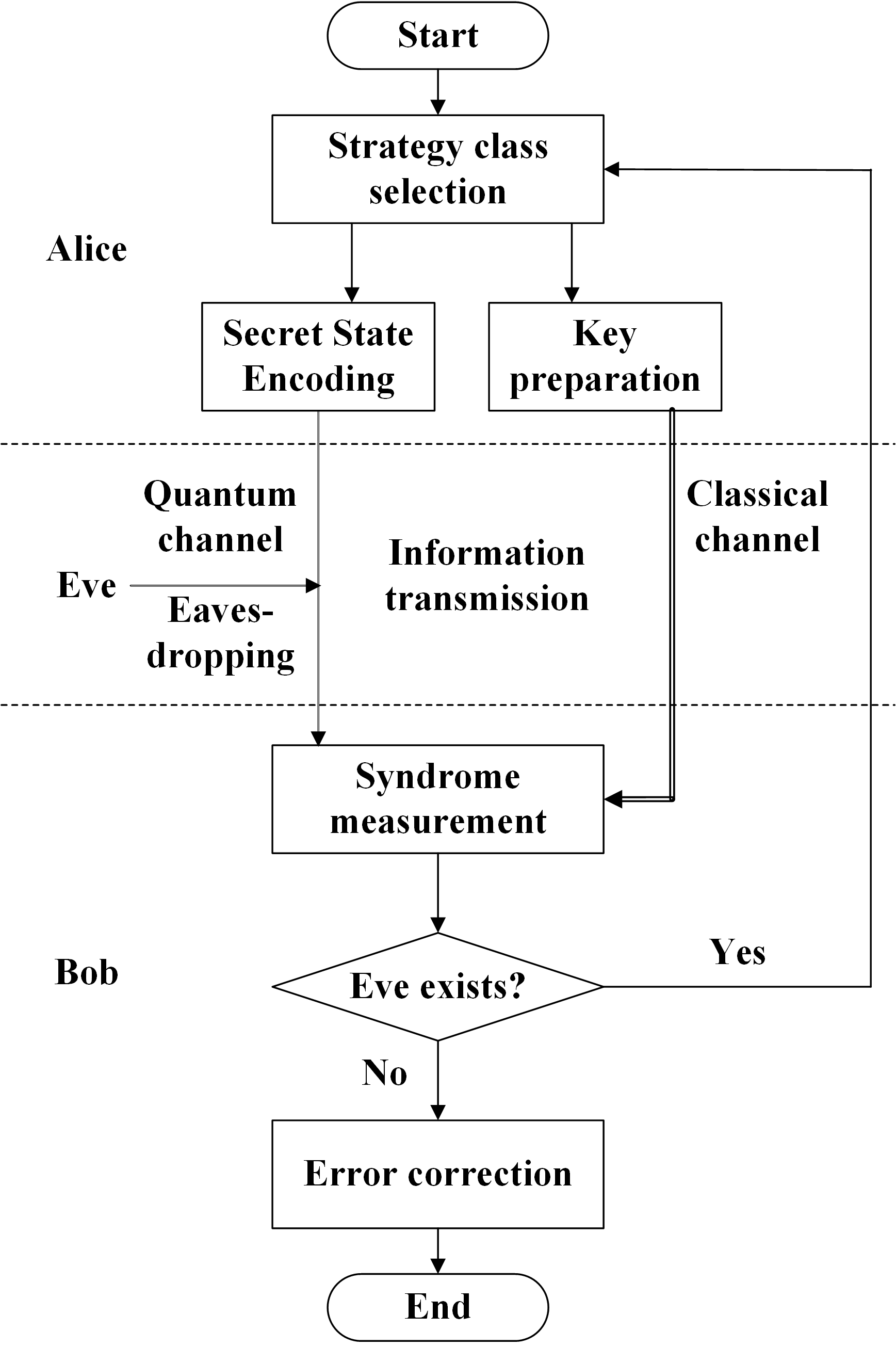}
\caption{Flowchart of the MGPD steganography protocol.}
\label{figure4}
\end{figure}

Drawing on the methods and strategies outlined in Sections \ref{section3.1} and \ref{section3.2}, we now provide a comprehensive description of the steganography protocol, as depicted in Fig. \ref{figure4}.

\textbf{Step1 (Transmitter: Strategy class selection)} Alice selects the class of strategy for modifying the projection direction based on the quantum system's channel environment. As detailed in Sect. \ref{section3.1}, for the bit-flip channel, the strategy class is  $\left\{\text{Original};g_4;g_5;g_6;g_4,g_5;g_4,g_6;g_5,g_6;g_4,g_5,g_6 \right\}$; for the phase-flip channel, it is \\ $\left\{\text{Original};g_1;g_2;g_3;g_1,g_2;g_1,g_3;g_2,g_3;g_1,g_2,g_3 \right\}$; and for the bit-phase-flip channel, it is \\ $\left\{\text{Original};g_1,g_4;g_2,g_5;g_3,g_6;g_1,g_2,g_4,g_5;g_1,g_3,g_4,g_6;g_2,g_3,g_5,g_6;g_1,g_2,g_3,g_4,g_5,g_6\right\}$.

\textbf{Step2 (Transmitter: Key preparation)} Alice employs the method outlined in Sect. \ref{section3.2}, leveraging her knowledge of the channel's error probability $p$, to determine the distortion parameter $\delta$, calculate the adjustment probability $p_g$, and construct the key sequence $K$ accordingly.

\textbf{Step3 (Transmitter: Secret state encoding)} Based on the intended secret message, whether classical or an arbitrary quantum state, Alice utilizes the key sequence $K$ and the strategy set chosen in Step1 to apply stabilizer encoding with modified generator directions, as shown in Fig. \ref{figure3}. Thus, the code stream carrying the secret information is generated, comprising logical encoded blocks, each consisting of 7 physical qubits.

\textbf{Step4 (Information transmission)} Alice transmits the carrier state stream $(\ket{\psi_{s1}},\ket{\psi_{s2}},...)$ through the quantum channel to Bob, while the key sequence $K$ and strategy set selected in Step1 are sent via a classical secure channel.

\textbf{Step 5 (Receiver: Syndrome measurement)} Upon receiving the erroneous quantum states $(\ket{\psi_{s1}^{'}},\ket{\psi_{s2}^{'}},...)$ through the channel, Bob measures them according to the generator projection directions corresponding to the strategy set rules specified by the key $K$ to obtain the error syndromes.

\textbf{Step 6 (Receiver: Eavesdropping detection)} Bob integrates the measurement outcomes from Step 5 and, guided by the key sequence $K$, determines an appropriate detection length. He conducts statistical analysis on every $k$-th position within each segment of length $\text{len}(K)$ in the state stream, (i.e. $\left\{\ket{\psi_{st}}\ |\ t\ \text{mod}\ \text{len}(K)=k\right\}$, corresponding to states subjected to the same MGPD), employing the method outlined in Sect. \ref{section3.1}, to ascertain the presence of eavesdropper Eve.

\textbf{Step 7 (Receiver: Error correction)} Should eavesdropping be detected, Bob alerts Alice to halt any further communication and discards the quantum states received in the current round. If no eavesdropping is detected, errors are rectified using the error correction module illustrated in Fig. \ref{figure3}. After discarding the irrelevant states at non-embeddable positions as indicated by key $K$, the recovered secret states can undergo post-processing and application.

\section{\label{section4}Analysis and Comparison with Related Work}

The security, imperceptibility, capacity, and robustness of the steganography framework proposed in this paper have been discussed in detail in the previous sections. This subsection provides a concise review of these characteristics and contrasts our scheme with prior steganography efforts based on QECC. Additionally, we juxtapose the protocol framework of our scheme with milestone works in QKD and QSDC from the perspective of confidential communication.

\subsection{\label{section4.1}Analysis}

\textbf{1) Security}: As detailed in Sect. \ref{section3.1}, if Eve intercepts the quantum states encoded with the $(7,1,3)$ code that contain the secret information, she can only perceive false syndromes, indicating incorrect error occurrences and locations (refer to Table \ref{table3}), unless she is aware of the actual generator projection directions modified by Alice. This results in erroneous error correction and post-processing, thereby ensuring the secrecy of the quantum states' content.

\textbf{2) Imperceptibility}: As outlined in Sect. \ref{section3.2}, Alice and Bob maintain the uniformity of the single physical bit error probability observed by Eve through the prearranged 'Sudoku' table (Table \ref{table6}). By managing the frequency of the MGPD with a shared key, they keep the deviation between the physical qubit error rate and the actual channel error rate within a small fluctuation $\delta$, making it challenging for Eve to discern.

\textbf{3) Capacity}: The capacity of steganography is typically denoted by the steganographic rate $r$ , the ratio of the number of secret messages to the number of carriers. As mentioned in Sect. \ref{section3.2}, Alice embeds secret quantum states at positions corresponding to the MGPD with a probability of $7p_g$. For the $(7,1,3)$ code, the steganographic rate is $r=(1/7)\times7p_g=p_g$. Moreover, capacity and imperceptibility are inversely related.

\textbf{4) Robustness}: The steganography scheme, which relies on stabilizer codes, inherently exhibits robustness against channel noise and errors. As discussed in Sect. \ref{section3.1}, while this scheme is not impervious to measure-resend attacks, it can detect the presence of the eavesdropper Eve, thus ensuring the security of the communication or verifying the reliability of the received quantum state data.

\textbf{5) Key Consumption}: According to the analysis in Sect. \ref{section3.2}, the length of the auxiliary key sequence required for the steganography scheme is a constant $C$, independent of the secret state sequence length, rendering this resource consumption negligible.

\subsection{\label{section4.2}Comparison}

The analysis presented above demonstrates that the steganography scheme outlined in this paper is applicable to both information hiding and quantum communication contexts. We now proceed to compare our scheme with selected prior works. Despite variations in form, framework, and procedure among these methods, we can identify common foundational criteria to facilitate a consistent and fair evaluation.

\subsubsection{\label{section4.2.1}Information Hiding Scenarios}

In the following, we contrast our MGPD scheme with other schemes as presented in Ref. \cite{ref12} and in Refs. \cite{ref13,ref14,ref15,ref16,ref17}, all of which are predicated on quantum error correction code-based information hiding techniques. As introduced earlier, the core concept of these schemes is to incorporate secret messages into the quantum error correction codes as artificial errors, to be later extracted using a key. Their specific implementations, however, diverge: Ref. \cite{ref12} employs a random selection of basis states and error directions ($X$ or $Z$), creating superpositions of various physical bit errors to detect any interference by Eve with the data. Differently, the schemes detailed in Refs. \cite{ref13,ref14,ref15,ref16,ref17}, embed secret information into the ancillary qubits of the QECC, managing the embedding rate to ensure that the error rate shows only a minor deviation from the channel's inherent characteristics. In contrast, our MGPD scheme forgoes the artificial embedding of errors corresponding secret message, instead integrating directional control during encoding and probabilistic direction selection adjustments.

Table \ref{table7} presents a comparison of the functionalities and performance metrics of the three types of schemes across various aspects. Here, $N$ denotes the total length of the carrier qubits, the key consumption rate $r_k$ is defined as the ratio of classical key number $n_k$ to the carrier length $N$, and $h(q)\equiv-(1-q)\log{(1-q)}-q\log{(q)}$ represents the channel entropy. All other basic settings and symbols are consistent with those previously described.

\begin{table}
\centering
\caption{Comparison of 3 types of quantum information hiding schemes.}
\begin{tabular}{p{0.14\linewidth}|p{0.17\linewidth}|p{0.2\linewidth}|p{0.25\linewidth}|p{0.18\linewidth}}

\toprule%第一道横线
\textbf{Scheme}& \textbf{Function}& \textbf{Channel form} & \textbf{Secret message form} & \textbf{Imperceptibility} \\
 \midrule%第二道横线
MGPD& Steganography& Single type of BC /PC /B-PC & Arbitrary quantum state/ Classical message & Strong \\  

Refs. \cite{ref13,ref14,ref15,ref16,ref17}& Steganography & Arbitrary type of BC/ PC /B-PC/ DC & Fixed form of quantum state/ Classical message & Strong \\

Refs. \cite{ref12}& Watermarking& Arbitrary type of BC/ PC /B-PC/ DC & Quantum superposition state & Weak\\
 \midrule%第二道横线
\textbf{Scheme}& \textbf{Resistance to channel noise}& \textbf{Eavesdropping Detection} & \textbf{Secret message embedding rate / Protected quantum data ratio} & \textbf{Key consumption} \\
 \midrule%第二道横线
MGPD&  $\surd$ & $\surd$ & $ r\leq \text{min}\left( \frac{\delta}{|1-8p|},\frac{1}{7}\right)$ & $n_k=O(1)$ \\

Refs. \cite{ref13,ref14,ref15,ref16,ref17}&  $\surd$ & $\times$ & $ r\leq \frac{4}{7}\delta\left(1-\frac{4}{3}p\right)$ & $n_k\approx O(\log{(N)})$ \\
Refs. \cite{ref12}&  $\surd$ & $\surd$ & $ r\leq \frac{1}{n}=\frac{1}{7}$ & $n_k=O(N)$ \\
\bottomrule%第三道横线
\end{tabular}%
\label{table7}%
\end{table}%

For the scheme presented in Refs. \cite{ref13,ref14,ref15,ref16,ref17}, the key consumption $n_k\approx \log{\left(\text{C}\left(N,\frac{\delta}{7}(1-\frac{4}{3}p)N\right)\right)}$. This represents a theoretical value, as the computation of such a combinatorial index and the retrieval of an element of length $N$ from a large dataset are, in practice, exceedingly demanding tasks. In the scheme in Ref. \cite{ref12}, $n_k=1/7N$. For large block size $n$ and carrier length $N$, the secret message embedding rate $r\approx h(p+\delta)-h(p)$, and the key consumption rate  tends toward $r_k\rightarrow h\left({\delta}/(1-2p)\right)-h(p+\delta)+h(p)$. Moreover, Reference \cite{ref12} proposes a theoretical method for data and message reversal, potentially increasing the ratio of protected quantum data to $r\leq(n-1)/n=6/7$. (Note: Due to the unification of basic conditions, some computational results differ in form from those in the original literature, but the methods remain consistent with the original sources.)

Examination of Table \ref{table7} reveals that, aside from its limitation to addressing single-type Pauli channel errors, our MGPD scheme exhibits several advantages. These include the ability to transmit arbitrary quantum states, a high degree of imperceptibility, and the capability for eavesdropping detection.

\begin{figure}
\centering
\includegraphics[width=0.7\textwidth]{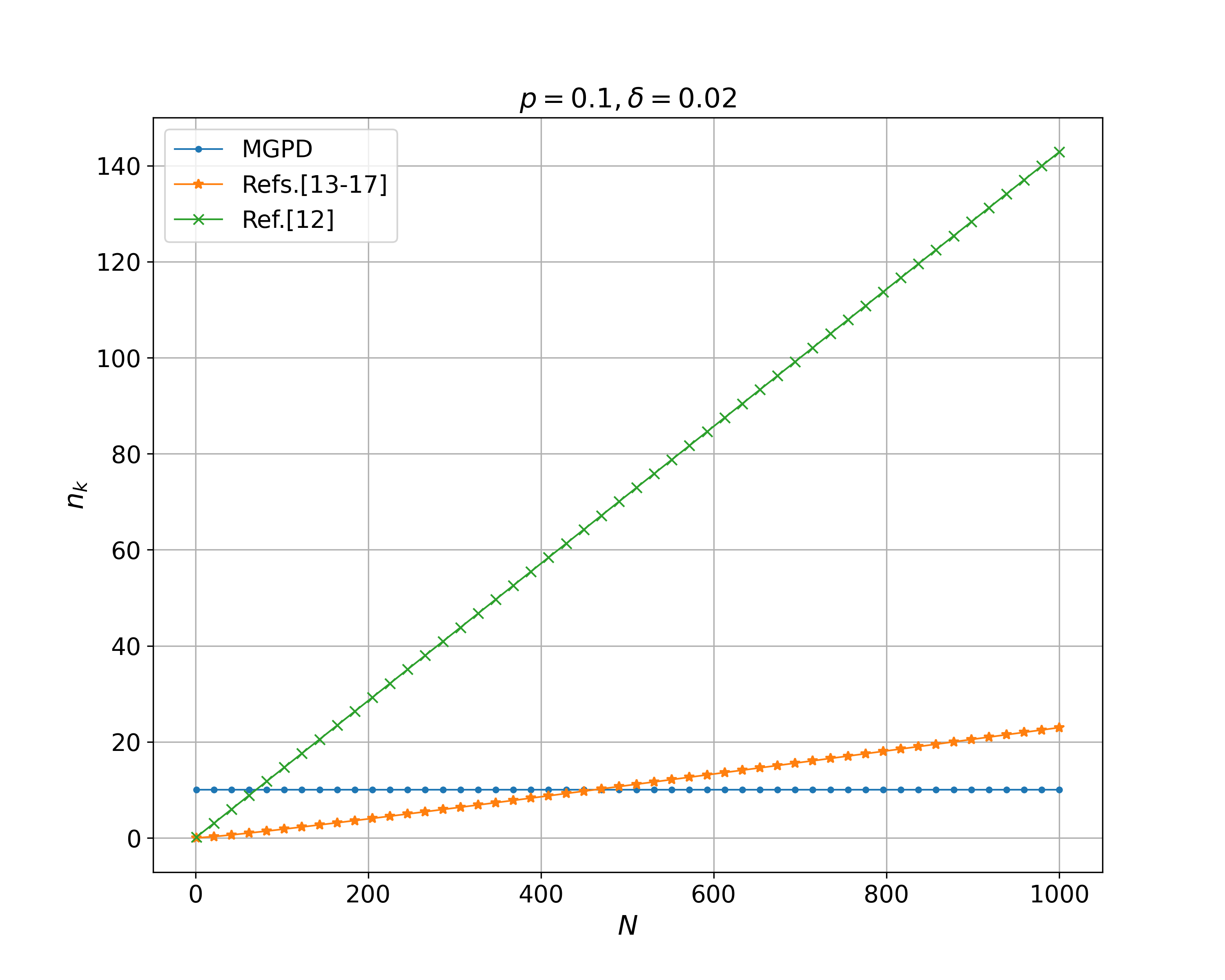}
\caption{Comparison of key consumption for 3 schemes. Here, the conditions are set to $p=0.1,\delta=0.02$, consistent with the example in Sect. \ref{section3.2}. The blue, orange, and green curves correspond to schemes MGPD, Refs. \cite{ref13,ref14,ref15,ref16,ref17}, and Ref. \cite{ref12}, respectively.}
\label{figure5}
\end{figure}

Significantly, the MGPD scheme necessitates only a constant-level key sequence, which, in comparison to the other two schemes, reduces key consumption from $O(\log{(N)})$  to $O(1)$ (refer to Fig. \ref{figure5}). This reduction decreases the overhead associated with auxiliary channels for key transmission and bolsters overall security. Since both schemes presented in Refs. \cite{ref13,ref14,ref15,ref16,ref17} and Ref. \cite{ref12} require the insertion of artificial errors to embed secret messages, compelling Alice and Bob to utilize random keys of comparable length to designate the positions and types of these embedded errors; otherwise, they would be vulnerable to detection and decryption by Eve. In contrast, the key in the MGPD scheme serves solely to align Alice and Bob on the encoding-measurement directions and does not involve the embedding of artificial errors. Errors within the channel occur randomly and are independent of the key, preventing Eve from making direct correlations.

\begin{figure}
\centering
\includegraphics[width=1\textwidth]{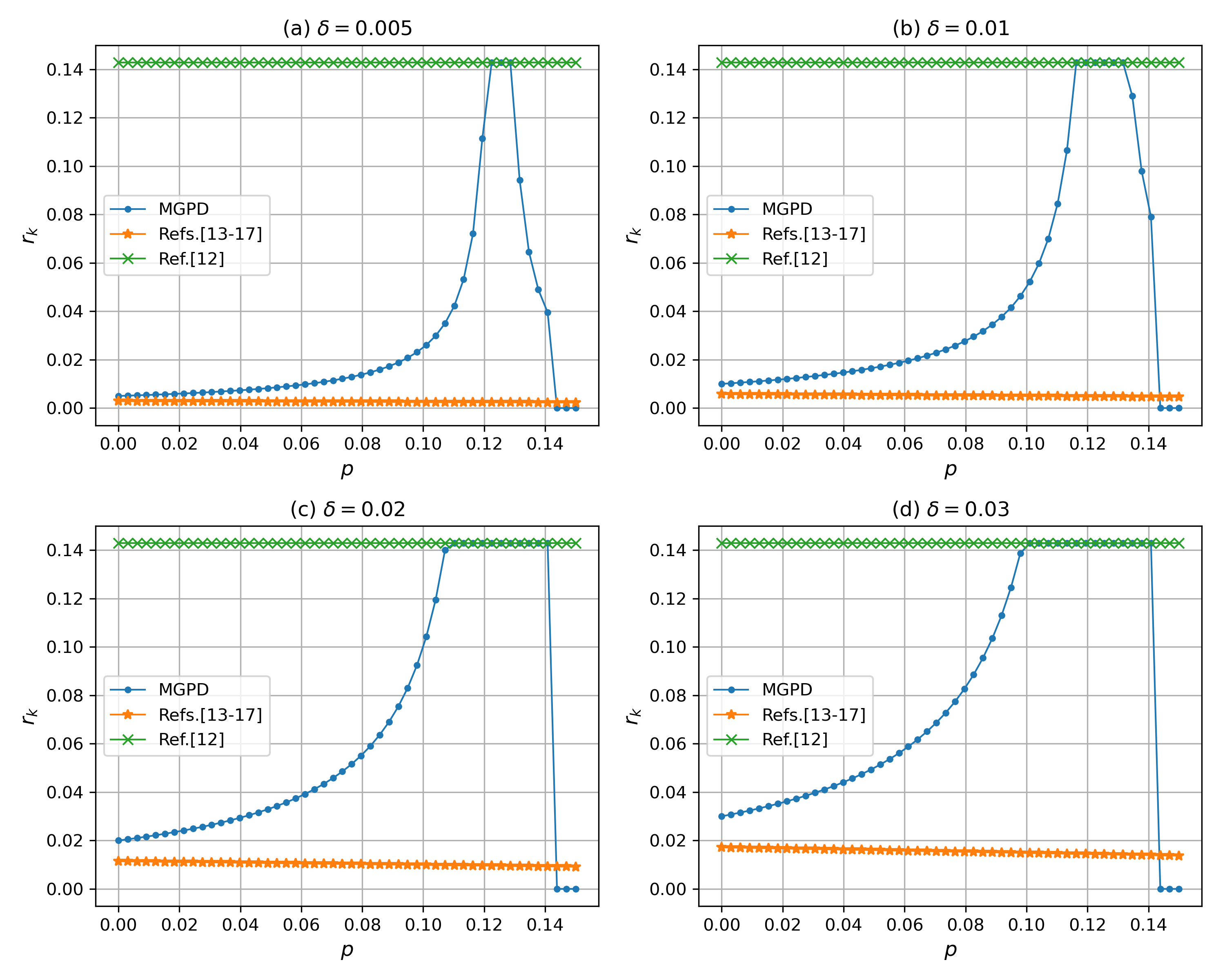}
\caption{Comparison of steganographic embedding rates for 3 schemes. Subfigures (a)-(d) correspond to deviation $\delta =0.005,0.01,0.02,0.03$. The blue, orange, and green curves correspond to schemes MGPD, Refs. \cite{ref13,ref14,ref15,ref16,ref17}, and Ref. \cite{ref12}, respectively.}
\label{figure6}
\end{figure}

Regarding the embedding rate, the MGPD scheme surpasses the methods presented in Refs. \cite{ref13,ref14,ref15,ref16,ref17} for $p\leq 1/7$, as shown in Fig. \ref{figure6}. Moreover, from Fig. \ref{figure6}, it can be observed that near $p= 1/8=1/(n+1)$, the embedding rate of the MGPD scheme approaches and attains the upper limit of the information-carrying capacity of the Steane code. This situation corresponds to the channel error probability being almost equal to the error-free probability (i.e., $p(I)=p(Z_k),k=1,...,7$). At this juncture, it becomes challenging for Eve to discern the statistical variations induced by the steganography, thereby achieving the maximum embedding rate. It should be noted that $p>1/7$ is not within the basic setting of the MGPD scheme described in this paper, and thus the embedding rate is set to zero under such conditions. Furthermore, as discussed at the end of Sect. \ref{section3.2}, an increased deviation setting for $\delta$ leads to a higher embedding rate, a phenomenon also illustrated in Fig. \ref{figure6}.

\subsubsection{\label{section4.2.2}Secure Communication Scenarios}
As previously discussed, the MGPD scheme is capable of transmitting messages in arbitrary quantum state form and offers features such as eavesdropping detection, resistance to Eve's analysis, and robustness against channel noise. A pertinent question emerges: Is this scheme applicable to QKD and QSDC scenarios? In this sections, we will engage in theoretical explorations and compare our scheme with seminal works in the fields of QKD and QSDC. Seminal works in QKD include the BB84 protocol \cite{ref4} and its entangled counterpart, the BBM92 protocol \cite{ref5}. These protocols capitalize on the uncertainty principle and the no-cloning theorem of quantum mechanics, exploiting Eve's lack of knowledge about the measurement bases and interactions over classical public channels to generate secure keys across quantum channels. In contrast, QSDC schemes \cite{ref8,ref9} do not necessitate the establishment of a shared key between the communicating parties. They encode messages onto non-local entangled states, necessitating only a minimal amount of classical information for eavesdropping detection or parameter estimation, thereby facilitating the direct transmission of secret messages over quantum channels.

In the task of quantum communication, there are several key performance metrics: 1) Key (or secret message) generation rate $R$ : This metric denotes the average amount of final secure information produced per initial qubit, with higher $R$ signifying greater efficiency. 2) The classical bit consumption rate $r_k$: This indicates the average quantity of auxiliary classical information needed to process one qubit, where a lower $r_k$ suggests enhanced efficiency and security. 3) Entanglement resources: This metric gauges the practical difficulty of implementing the scheme, with higher entanglement resource demands posing greater challenges for practical application. 4) KL Divergence $D_{KL}$ before and after eavesdropping: This quantifies the alteration in error probability distribution due to eavesdropping, computed as $D_{\text{KL}}\equiv\sum_x P(x)\log{\left(P(x)/Q(x)\right)}$ \cite{ref32}. A higher $D_{KL}$ denotes a more robust eavesdropping detection capability. 5) The capacity to transmit the form of Q-messages: In scenarios necessitating the direct presentation of quantum state data, the quantum state form of secret information holds significant potential for applications in information security. 

In the majority of studies within the realms of QKD and QSDC, a thorough analysis of channel noise has not been conducted. In the following section, we present a comparison of the functionalities and performance metrics of the three classes of schemes under a channel noise probability $p$, as depicted in Table \ref{table8}. Here $\varepsilon$ signifies the proportion of qubits dedicated to the eavesdropping detection process, which can be quite minimal. In experiments from \cite{ref33}, $\varepsilon$ achieved good detection results at 3\% and nearly 100\% accuracy at 20\%. Other settings remain consistent with the previous discussion.

\begin{table}
\centering
\caption{Comparison of 3 types of quantum information hiding schemes.}
\begin{tabular}{p{0.1\linewidth}|p{0.15\linewidth}|p{0.15\linewidth}|p{0.15\linewidth}|p{0.16\linewidth}|p{0.22\linewidth}}

\toprule%第一道横线
\textbf{Scheme}& \textbf{Form of secret messages}& \textbf{Key generation rate} & \textbf{Classical bit consumption rate} & \textbf{Entanglement resources required} & \textbf{KL Divergence before and after eavesdropping} \\
 \midrule%第二道横线
MGPD & arbitrary qubits $\alpha\ket{0}+\beta\ket{1}$ or classical bits & $R=\text{min}\left(\frac{\delta}{|1-8p|},\frac{1}{7}\right)$ & $r_k\ll1 $ & QECC, short-distance & $D_{\text{KL}}=p\log{\frac{p}{1-7p}}+(1-7p)\log{\frac{1-7p}{p}}$\\  
 \midrule%第二道横线
BB84 \cite{ref4}& 0,1 bits & $R\approx\frac{1-\varepsilon}{2}\left(1-h(p)\right)$ & $r_k\approx2+\varepsilon$ & don’t require & $D_{\text{KL}}=p\log{\frac{p}{1/4+p-p^2}}+(1-p)\log{\frac{1-p}{3/4+p^2-p}}$\\
 \midrule%第二道横线
Two-step QSDC \cite{ref9}& 0,1 bits or Bell states& $R\approx(1-2\varepsilon)\left(1-h(p)\right)$ & $r_k\approx2\varepsilon$ & EPR pairs, long-distance & 1st detection: same as BB84; 2nd detection: $D_{\text{KL}}=(1-p)\log{4(1-p)}+p\log{\frac{4}{3}p}$\\

\bottomrule%第三道横线
\end{tabular}%
\label{table8}%
\end{table}%

As shown in Table \ref{table8}, both the QSDC and MGPD schemes exhibit low auxiliary classical bit consumption rates, underscoring the benefits of 'direct' communication via quantum channels. Regarding entanglement resources, the BB84 protocol is the most straightforward to implement, as it does not necessitate entanglement. The MGPD protocol relies on local entanglement within QECC blocks, posing a moderate implementation challenge. Conversely, the QSDC protocol involves the distribution of entangled particles over extended distances, which is most prone to decoherence and incurs the highest implementation costs. In terms of the form of secret information, the MGPD scheme is capable of transmitting arbitrary quantum states, including classical 0- and 1-bits represented by basis states, thus offering a wider range of potential applications compared to the other two schemes.

\begin{figure}
\centering
\includegraphics[width=0.6\textwidth]{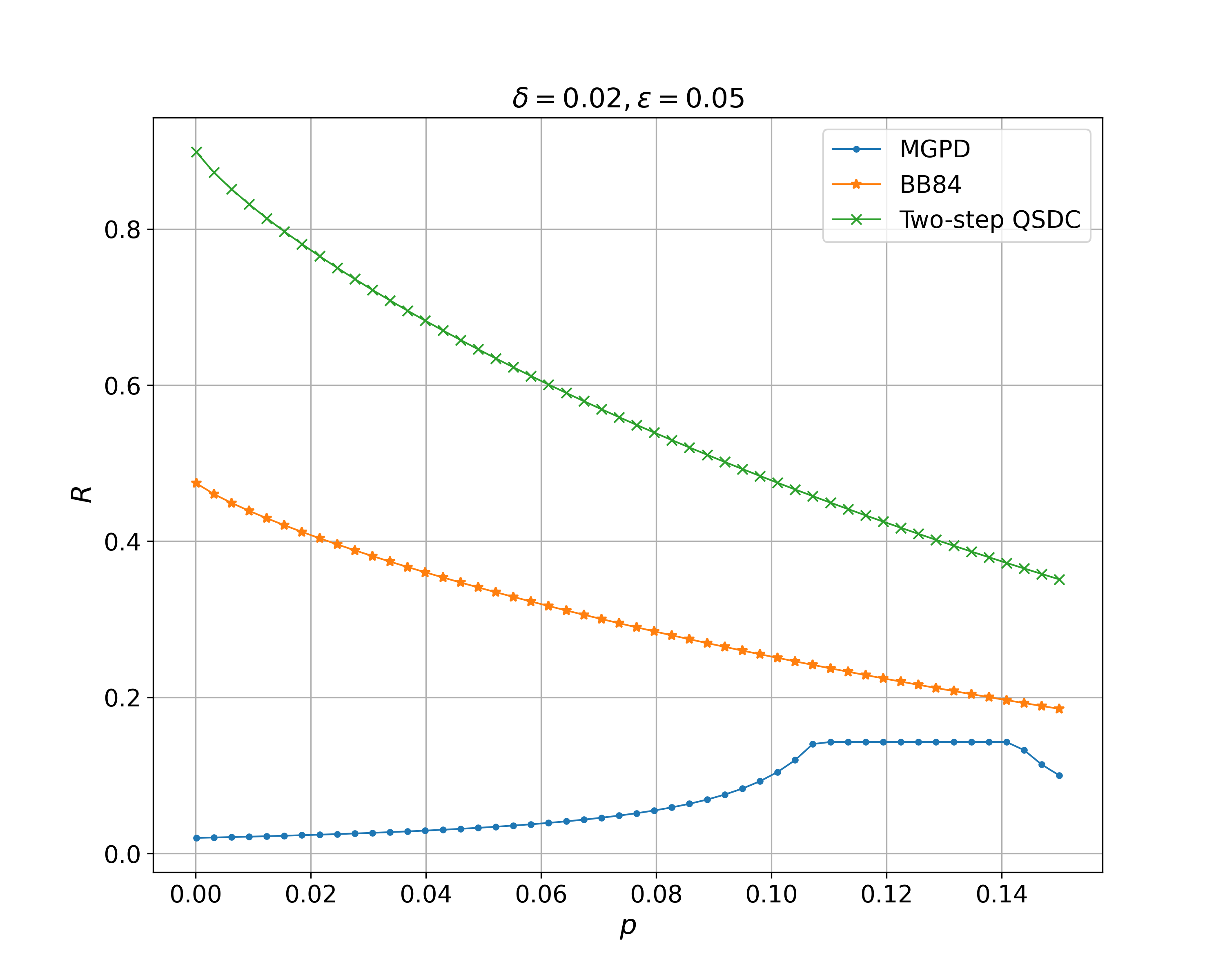}
\caption{Comparison of key generation rate for 3 schemes. Here, the conditions are set to $\delta=0.02$ and $\varepsilon=0.05$. The blue, orange, and green curves correspond to schemes MGPD, BB84, and two-step QSDC, respectively.}
\label{figure7}
\end{figure}

Figure \ref{figure7} illustrates the key generation rate $R$ as a function of the channel noise probability $p$. It is observable that the MGPD scheme does not possess a capacity advantage when the channel error probability $p$ is exceedingly low. Within the range $0.1 \leq p\leq0.14$, its capacity approximates and attains the upper limit of the information capacity of the Steane code, showing no significant difference from the BB84 protocol.

\begin{figure}
\centering
\includegraphics[width=1\textwidth]{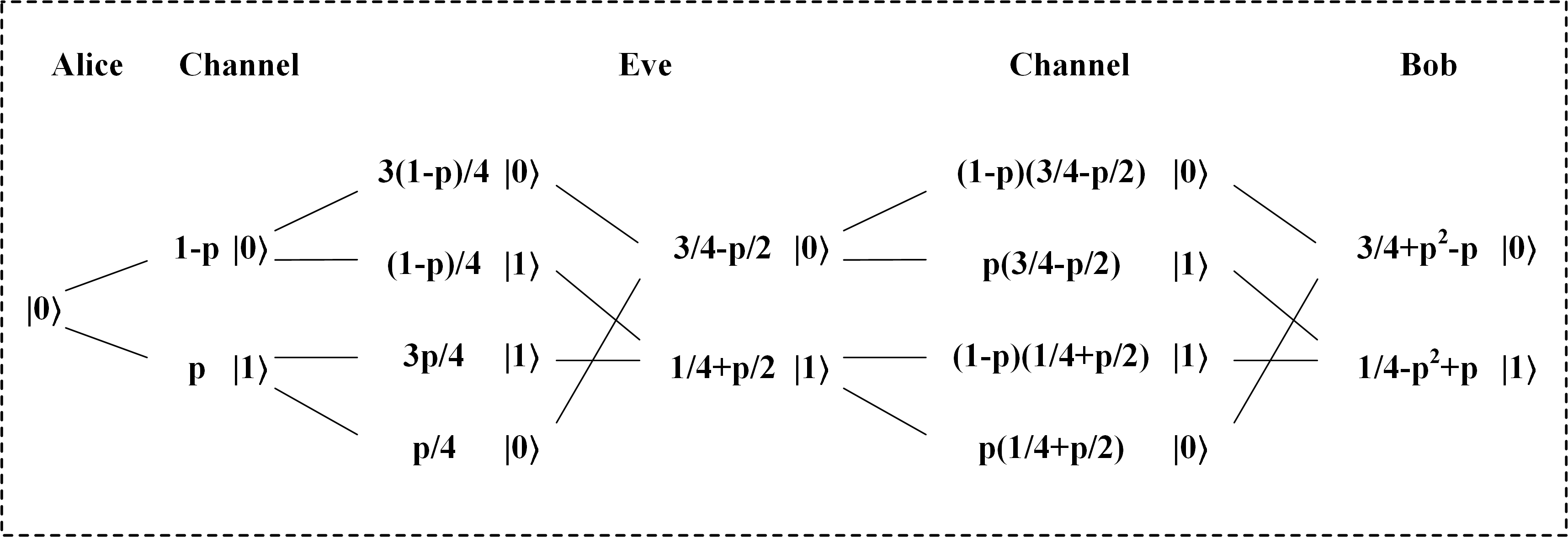}
\caption{Schematic diagram illustrating the error rate calculation process for the BB84 scheme under eavesdropping. The figure uses the transmission of the $\ket{0}$ state, $\ket{1}$ state follows the same process. Note that in the final step, Bob will compare his measurement basis with Alice, ensuring that there are no errors due to incorrect basis selection.}
\label{figure8}
\end{figure}

Regarding KL divergence, in the context of the BB84 protocol, Eve's intercept-resend attack results in qubits being subjected to the effects of eavesdropping measurements as well as two rounds of channel noise errors. The total error rate is depicted in Figure \ref{figure8}. By integrating this with the error rate $p$ in the absence of eavesdropping, we can derive the results presented in Table \ref{table8}.

In the Two-Step QSDC scheme, the first eavesdropping detection adheres to the principle analogous to that of the BB84 protocol, yielding an identical $D_{KL}$. In the second detection phase, Eve's intercept-resend attack disrupts the EPR pairs shared between Alice and Bob, causing Bob to receive a maximally mixed state $I/4$. Consequently, Bob's measurement outcomes will be one of the four Bell states with equal likelihood, resulting in an error probability of $3/4$. This accounts for the results detailed in Table \ref{table8}.

\begin{figure}
\centering
\includegraphics[width=0.6\textwidth]{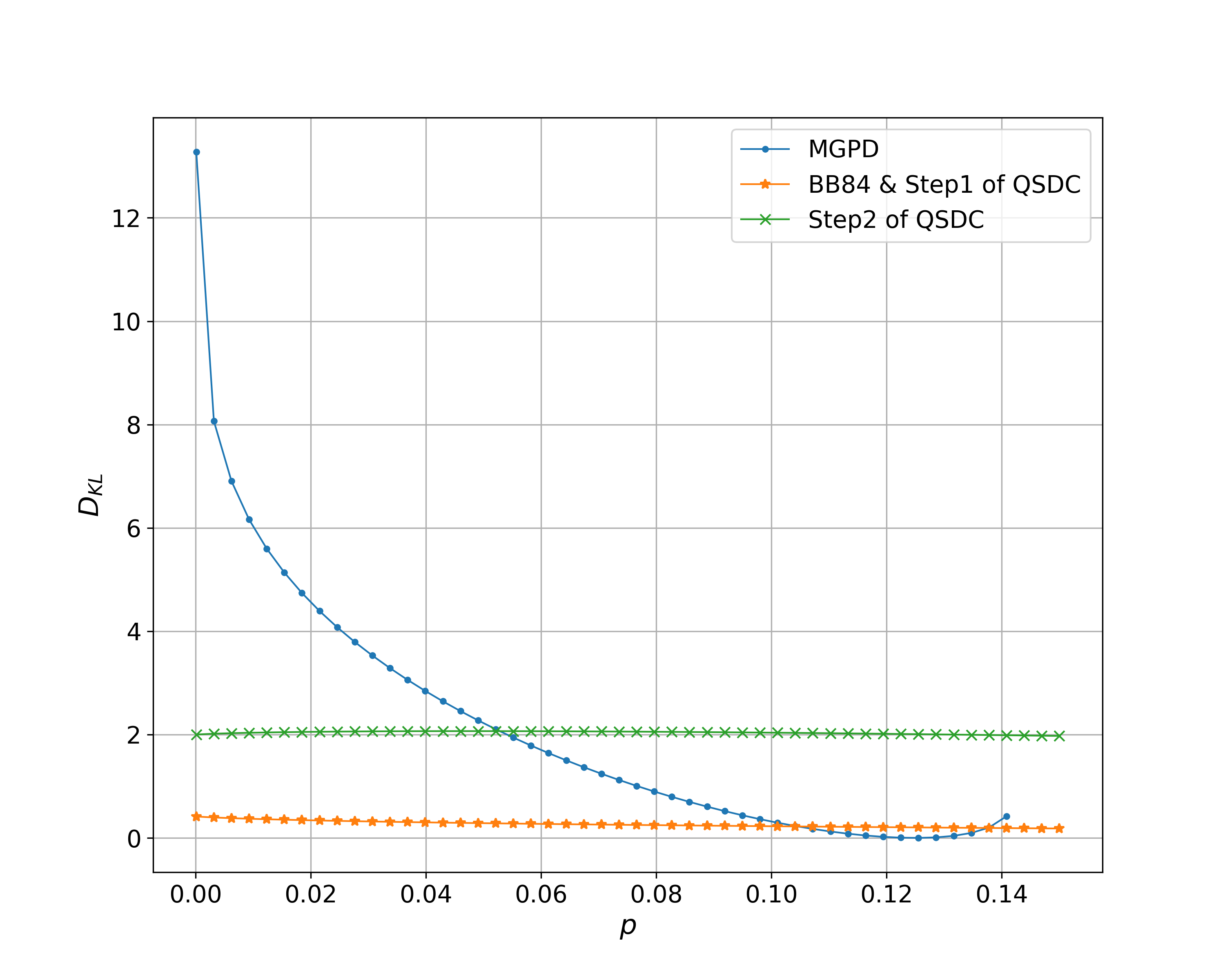}
\caption{Comparison of KL divergence before and after eavesdropping for 3 schemes. The blue, orange, and green curves correspond to schemes MGPD, BB84 \& step1 of QSDC, and step2 of QSDC, respectively.}
\label{figure9}
\end{figure}

The KL divergence of the three schemes as a function of the noise probability $p$ is shown in Fig. \ref{figure9}. It can be seen that at  $p\leq0.05$, the MGPD scheme has the strongest eavesdropping detection capability. At $0.05<p<0.14$, it also performs overall better than the BB84 scheme. Of course, as explained at the end of Sect. \ref{section3.2}, there is a trade-off between the two indicators of security and capacity with the variation of $p$.

In summary, we have investigated the potential applications of the MGPD scheme in the realm of quantum secure communication. While the concept is idealized, the scheme's capacity to transmit arbitrary quantum states, coupled with its eavesdropping detection and immediate error correction capabilities, holds potential promise in specific applications, such as quantum state-based key distribution, sharing, and the establishment of quantum networks.

\section{\label{section5}Conclusion}
This study presents a quantum steganography method designed for the direct transmission of secret quantum states. Distinct from quantum cryptography, quantum steganography utilizes the intrinsic properties of quantum channels and quantum information encoding to ensure that eavesdroppers cannot discern the presence of covert information. Though this domain is nascent, it possesses substantial potential for future quantum communication and networking, especially in covert communication, identity authentication, and tamper detection. Moreover, it can be integrated with quantum cryptography to construct more robust information security frameworks.

We introduce a steganographic encoding method based on the Steane code, which involves modifying the projection directions of the generators. We theoretically demonstrate that an eavesdropper is incapable of correctly decoding the information. We develop detailed encoding and decoding schemes, quantum circuits, and principles for eavesdropping detection, employing a 'Sudoku'-style strategy to balance error probabilities, and present the comprehensive steganography protocol. Compared to related research on quantum information hiding based on QECC, the MGPD scheme offers benefits in terms of secret information form, stealth, security, and capacity, notably by reducing auxiliary key consumption from $O(\log{(N)})$ to $O(1)$. We also explore the MGPD scheme's potential for applications such as QKD and QSDC, underscoring its strengths in secret information form and eavesdropping detection within noisy channels.

The limitations of our work are primarily twofold: 1) We made idealized assumptions, focusing solely on single-type Pauli noise, which may not be applicable in scenarios involving multiple Pauli noises or depolarizing channels. 2) In contrast to QKD, which employs classical post-processing for error correction, QECC incurs higher quantum resource costs, leading to a lower secret information generation rate (with an upper bound of $1/n=1/7$ for the Steane code).

Moving forward, we aim to enhance the scheme's applicability to various noise channels and to improve the secret information generation rate. The quantum $(5,1,3)$ code, also known as the Laflamme code \cite{ref34}, is a promising QECC candidate for tackling these challenges. Additionally, investigating which $(n,k,d)$ stabilizer codes can be adapted for information steganography using the MGPD method presents an intriguing theoretical question.

\begin{acknowledgments}
This work is supported by the National Natural Science Foundation of China under Grant No. 62171470, Henan Province Central Plains Science and Technology Innovation Leading Talent Project (No. 234200510019), Natural Science  Foundation of Henan Province (No. 232300421240) and Laboratory for Advanced Computing and Intelligence Engineering Fund.
\end{acknowledgments}

\nocite{*}
% \section{\label{section6}References}
\bibliography{apssamp}% Produces the bibliography via BibTeX.

\end{document}